\documentclass[aps,prl,twocolumn]{revtex4}

\usepackage{natbib}
\usepackage{graphicx}
\usepackage{amssymb, amsmath}
\usepackage{todonotes}

\newcommand{\aap}{Astron.\ Astrophys.}
\newcommand{\mnras}{Mon.\ Not.\ R.\ Astron.\ Soc.}

\newcommand{\apjl}{Astrophys.\ J.\ Lett.}
\newcommand{\apjs}{Astrophys.\ J.\ Suppl.\ Ser.}

\def\la{\; \raise0.3ex\hbox{$<$\kern-0.75em\raise-1.1ex\hbox{$\sim$}}\;}
\def\ga{\;  \raise0.3ex\hbox{$>$\kern-0.75em\raise-1.1ex\hbox{$\sim$}}\;}

\usepackage[normalem]{ulem}

\begin{document}

\title{Constraining neutron superfluidity with $r$-mode physics}
\author{Elena M. Kantor}
\author{Mikhail E. Gusakov}
\author{Vasiliy A. Dommes}
\affiliation{ Ioffe Institute, Polytekhnicheskaya 26, 194021
St.-Petersburg, Russia }


\begin{abstract}
	
We constrain 
the parameters of neutron superfluidity in the cores of neutron stars
making use of the
recently proposed effect of resonance stabilization of $r$-modes. 
To this end, we, for the first time, 
calculate the finite-temperature $r$-mode spectra
for realistic models of rotating superfluid neutron stars,
accounting for both muons 
and neutron-proton entrainment in 
their
interiors.
We find that the ordinary (normal) $r$-mode exhibits avoided crossings 
with superfluid $r$-modes at certain 
stellar temperatures and spin frequencies.
Near 
the avoided crossings, the normal $r$-mode dissipates strongly, 
which leads to substantial suppression of the $r$-mode instability 
there.
The extreme sensitivity of the positions of avoided crossings 
to 
the neutron superfluidity model allows us to constrain 
the latter
by confronting 
the calculated spectra with observations 
of rapidly rotating neutron stars in low-mass X-ray binaries.
\end{abstract}


 \maketitle

\textit{Introduction.--}
Since 1998, when it was discovered \cite{andersson98,fm98}, 
the $r$-mode instability 
has been a problem 
for neutron star (NS) physics \cite{lom98,ltv01,rb03,mms08,hah11,as14a,gck14b,hp17}.
$R$-modes are 
predominantly toroidal oscillations 
(i.e., oscillations with a divergenceless velocity field and a suppressed radial component of the velocity \cite{ak01})
of rotating stars restored 
by the Coriolis force.
Their close relatives are 
Rossby waves on Earth.
Neglecting dissipation, they are unstable 
with respect to gravitational radiation
at any spin frequency $\nu$ of an NS. 
Dissipation, however, suppresses this instability to some extent. 
Cold and hot NSs are effectively stabilized 
by shear and bulk viscosities, respectively \cite{ak01}, 
but the stabilizing mechanism for rapidly rotating NSs of intermediate 
internal
temperatures, 
$T^\infty\sim 10^8\,\rm K$ 
(redshifted, as seen by a distant observer),
is not so obvious.   
Seemingly, we should not observe NSs in the region of 
$\nu$ and $T^\infty$
where the $r$-mode instability is not suppressed by dissipation (the ``instability window''),
since, as the theory predicts, 
the excited $r$-mode rapidly spins the star down by means of gravitational radiation 
so that the probability of finding an NS in this region 
is negligible \cite{levin99,gck14a,haskell15}.
However, the observations show \cite{hdh12,gck14a,cgk17} that numerous NSs fall well inside the classical 
(i.e., calculated under minimal assumptions) 
instability window. 
These stars belong to low-mass X-ray binaries (LMXBs), 
where they are heated and spun up by accretion from the low-mass companion. 
A series of proposals have been suggested 
to solve the puzzle of NSs in the instability window
(see \cite{haskell15,gg18} for the reviews). 
Most of them either involve exotic physics (e.g., a quark or hyperon composition
of the NS core) or 
make some model-dependent 
assumptions about the mechanism of the nonlinear saturation of $r$-modes \cite{haskell15}.
One of the 
proposals
is the resonance stabilization 
of $r$-mode by superfluid (SF) modes \cite{gck14a,gck14b,cgk14}.
The latter does not require any exotic physics and adopts standard 
assumptions about 
the properties of NS matter, the same as in the minimal cooling scenario 
\cite{plps04,gkyg04}.

\textit{Resonance $r$-mode stabilization scenario.--}
We consider the $npe\mu$-composition of an NS core: 
neutrons ($n$), protons ($p$), electrons ($e$), and muons  ($\mu$),
and account for a possible nucleon SF \cite{yls99,ls01}.
SF NSs can support several independent velocity fields: the velocity of SF neutrons and the velocity of the
remaining components \cite{khalatnikov89, ga06}
(the proton SF velocity is not independent 
since protons are coupled with electrons and muons
by electromagnetic forces).
As a result, new SF modes appear in the NS oscillation spectrum
in addition to ordinary (normal) modes of non-SF NSs \cite{lm94,lee95, gkgc14}.
In contrast to the normal modes, 
SF modes 
strongly depend on $T^\infty$ 
because of temperature dependence of neutron SF density \cite{khalatnikov89,ab76,bjk96,gh05,ch06,chamel08,gkh09b}
(more precisely, of the entrainment matrix $Y_{ik}$; see below).
Consequently, normal and SF modes exhibit avoided crossings 
at a certain (resonance) $T^\infty$, where the eigenfrequency of the SF mode approaches 
that 
of the normal mode. 
SF modes dissipate efficiently due to powerful mutual-friction mechanism \cite{als84}, 
which tends to equalize the velocities of normal and SF components, 
while for normal $r$-modes mutual friction is, generally, not effective.
However, at resonance $T^\infty$, the normal and SF modes interact strongly, 
the eigenfunction of the SF mode admixes 
to the eigenfunction of the normal mode, 
and the latter experiences resonance stabilization by mutual friction at these $T^\infty$ \cite{gck14a,gck14b}.
The scenario of \cite{gck14a,gck14b} 
uses this 
property to stabilize normal $r$-modes for NSs 
observed in the classical instability window.
According to 
\cite{gck14a,gck14b},
NSs in that window 
should have $\nu$ and $T^\infty$ corresponding to resonances between the SF and normal modes.

Initially, this scenario was proposed as a purely phenomenological one.
In this Letter, we
develop it into a quantitative theory.
To this end, 
we
calculate temperature-dependent $r$-mode spectra of rotating SF NSs
for realistic three-layer stellar configurations, consisting of a barotropic crust
treated as a single fluid, 
an npe
outer core and an npe$\mu$
inner core.
Calculations are performed adopting up-to-date microphysics input, including nonzero entrainment between neutrons and protons,
realistic equations of state (EOSs), and the parameters of a nucleon SF.
Confronting the calculated spectra with the available observations of NSs in LMXBs
allows us to put constraints on the neutron
SF critical temperature profile in the NS core.
 

\textit{Oscillation equations.--}
We consider nondissipative oscillations of a slowly rotating 
(spin frequency $\Omega=2\pi \nu$) NS. 
We adopt the Cowling approximation 
(i.e., neglect metric perturbations \cite{cowling41})
and 
work in the Newtonian framework. 
The linearized equations governing small oscillations of SF NSs 
in the frame rotating with the star are  as follows \cite{kg17}:

(i) Euler equation
\begin{eqnarray}
\frac{\partial {\pmb v}_{b}}{\partial t}+2 {\pmb \Omega}\times {{\pmb v}_{b}}=
-\delta\left(\frac{{\pmb \nabla} P}{w} \right)=
\frac{\delta w}{w^2}{\pmb \nabla} P-\frac{{\pmb \nabla} \delta P}{w}, 
\label{euler}
\end{eqnarray}
where $w=(P+\epsilon)/c^2$, $P$ is the pressure, $\epsilon$ is the energy density (including the rest mass energy density),
$c$ is the speed of light, and $t$ is time. 
Here and hereafter, 
$\delta$ stands for the Eulerian perturbation of the corresponding thermodynamic parameter.
The 
small (first order) perturbation of the velocity of baryons is
${\pmb v}_{b}\equiv {\pmb j}_{b}/n_{b}$,
where $n_{b}\equiv n_{n}+n_{p}$, ${\pmb j}_{b}\equiv {\pmb j}_{n}+{\pmb j}_{p}$, and $n_i$, ${\pmb j}_i$ ($i=n,p$) are the particle number density and current density, respectively.

(ii) Continuity equations for baryons and leptons 
(${l}=e, \, \mu$)
\begin{eqnarray}
\frac{\partial \delta n_{b}}{\partial t}+{\rm div}(n_{b} {\pmb v}_{b})=0 \label{cont b}, \quad
\frac{\partial \delta n_{l}}{\partial t}+{\rm div}(n_{l} {\pmb v})=0 \label{cont l},
\end{eqnarray}
where ${\pmb v}$
is the small (first order) perturbation of the velocity
of normal liquid component (leptons and baryon thermal excitations).

(iii) The ``superfluid'' equation 
[a combination of the Euler equation 
for the SF neutron liquid component and Eq.\ (\ref{euler})]
\begin{eqnarray}
-h \frac{\partial {\pmb v_\Delta}}{\partial t}- 2 h_1  {\pmb \Omega} \times {\pmb v_\Delta}=c^2 n_{e} {\pmb \nabla} \Delta \mu_{e}+c^2 n_{\mu} {\pmb \nabla} \Delta \mu_{\mu},
\label{sfl1}
\end{eqnarray}
where ${\pmb v_\Delta}\equiv{\pmb v}_{b}-{\pmb v}$;
$\Delta \mu_{l}\equiv \mu_{n}-\mu_{p}-\mu_{l}$ is the chemical potential imbalance;
in thermodynamic equilibrium $\Delta \mu_{l}=0$ \cite{hpy07}.
Further,
\begin{eqnarray}
h=n_{b} \mu_{n} \left[\frac{n_{b}Y_{{pp}}}{\mu_{n}(Y_{{nn}}Y_{{pp}}-Y_{{np}}^2)}-1\right], \label{beta} \\
h_1=n_{b} \mu_{n} \left(\frac{n_{b}}{Y_{nn}\mu_{n} + Y_{np}\mu_{p}}-1\right),
\label{gamma} 
\end{eqnarray}
where
$Y_{ik}$ ($i,k={n},p$) is the temperature-dependent 
entrainment matrix \cite{ab76,ga06,gkh09a,gkh09b,ghk14}, which is a
generalization of the concept of SF density 
(e.g., \cite{khalatnikov89}) 
to the case of SF mixtures.
The off-diagonal elements $Y_{np}=Y_{pn}$ 
of this matrix describe the ``entrainment effect'', i.e., 
the mass transfer of one SF particle species 
by the SF motion of another species.
Typically, 
this effect is rather weak 
(in other words, $Y_{np}$ is sufficiently small; see, e.g., \cite{dkg19}),
so that $h_1\approx h$.
Eqs.\ (\ref{euler})--(\ref{gamma}) 
should be supplemented with this 
relation:
$
\delta n_\alpha=\frac{\partial n_\alpha}{\partial P} \delta P+\frac{\partial n_\alpha}{\partial \Delta \mu_{e}} \Delta \mu_{e}+\frac{\partial n_\alpha}{\partial \Delta \mu_{\mu}} \Delta \mu_{\mu}$,
$\alpha=n,p,e,\mu$.

Now,
consider NS perturbations that depend
on time $t$ as ${\rm e}^{\imath\sigma t}$ 
in the corotating frame. 
Then, following the same procedure as for non-SF stars 
(e.g., \cite{lf99}), we expand all the unknown functions
into spherical harmonics ${\rm Y}_{lm}$ with fixed $m$.
In addition, we also expand all the quantities in a power series 
in small parameter $\Omega$ 
(here and in what follows we normalize $\Omega$ and $\sigma$ to the Kepler frequency).
We are interested in 
the oscillations, that are absent in nonrotating stars, 
i.e., that have the eigenfrequencies $\sigma$ 
vanishing at $\Omega \rightarrow 0$. 
In this case, $\sigma$ 
can be represented as follows (e.g., \cite{saio82,pbr81,lf99}):
$\sigma= \sigma_0 \Omega+O(\Omega^3)$.
When considering NSs with SF $npe\mu$ cores,
we found in 
\cite{kg17} that, for vanishing entrainment 
[when $Y_{np}=Y_{pn}=0$, i.e., $h_1=h$; see Eqs.\ (\ref{beta}), (\ref{gamma})], 
purely toroidal modes (in the lowest order in $\Omega$) 
are only possible if $l=m$. 
For a given $m$ there exist
one normal $r$-mode and an infinite set of SF $r$-modes,
all having the same $\sigma_0$:
$\sigma_0 = 2/(m+1)$
\cite{ac01,ly03,agh09,kg17}.

When neutron and proton SFs coexist somewhere in an NS, 
entrainment between neutrons and protons,
although weak, plays 
an important role, because it affects the position of avoided crossings as discussed below.
Assuming that the entrainment effect is 
small, 
we develop a
perturbation theory in the parameter 
$\Delta h \equiv h/h_1 - 1\ll 1$ \cite{dkg19}.
We account for both the corrections due to entrainment 
and next-to-leading order corrections in $\Omega$
in oscillation equations,
and 
treat them simultaneously, expanding the oscillation frequency as
\begin{eqnarray}
	\sigma 
	= (\sigma_{0}  + \sigma_{1})\Omega,
	\label{expan}
\end{eqnarray}
where 
$\sigma_{1}$
corresponds to the next-to-leading-order correction in $\Omega$ and $\Delta h$. 

\begin{figure}
    \begin{center}
        \leavevmode
        \includegraphics[width=8.5cm]{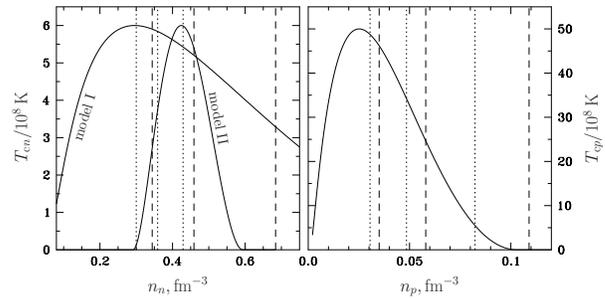}
    \end{center}
    \caption{
Profiles $T_{{\rm c}n}(n_{n})$ and $T_{{\rm c}p}(n_{p})$, respectively, for the neutron SF models I and II (left panel) and 
the proton SF model (right panel).
Vertical lines (dashes for APR EOS, dots for BSk24 EOS) show the central number densities of NSs with, 
from left to right, $M=1.0M_\odot,\,1.4M_\odot$, and $1.8M_\odot$.
		}
    \label{Fig:Tc}
\end{figure}

\textit{Physics input.--}
We consider two realistic EOSs. 
The first one (denoted as APR EOS) uses the parametrization \cite{hh99} of Akmal-Pandharipande-Ravenhall (APR) EOS from 
\cite{apr98}, 
and adopts $Y_{ik}$ from 
\cite{gh05}. 
The second one (BSk24 EOS) is obtained from the BSk24 energy-density functional
\cite{gcp13,pcp18};
the same functional is used to calculate $Y_{ik}$ in a self-consistent manner 
 \cite{ch06,gkh09a,gkh09b,ghk14}.
For each of these EOSs, we consider 
one model of a proton SF and two models, I and II, of neutron SF,
that differ by the width of the density-dependent profile of the (local)
neutron critical temperature, $T_{{\rm c}n}(n_n)$ (Fig.\ \ref{Fig:Tc}).
The adopted models have maximum critical temperatures that do not contradict 
the existing data on cooling NSs \cite{gkyg04,plps04,gkyg05,syhhp11,page11,CasA13,CasA15,bhsp18}
and the microscopic calculations \cite{ls01, yls99,gps14,dlz14, drddwcp16, sc19}.
The wider profile of $T_{{\rm c}n}$ (model I), 
which extends to lower densities, 
is favored more 
by the microscopic theory.
In turn, the narrower profiles (similar to model II) 
have been used
in a number of works 
(e.g., \cite{gkyg04,gkyg05,syhhp11,CasA13})
to successfully explain the thermal properties 
of isolated NSs within the minimal cooling scenario \cite{plps04,gkyg04}.

\begin{figure}
    \begin{center}
        \leavevmode
        \includegraphics[width=6cm]{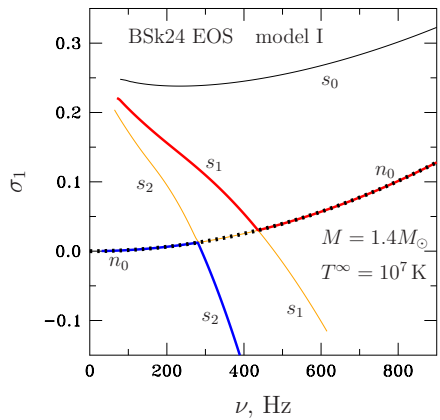}
    \end{center}
    \caption{$\sigma_1$ versus $\nu$
    	for normal $r$-mode ($n_0$) and first three SF $r$-modes ($s_0$, $s_1$, $s_2$).
    }
    \label{Fig:sigma1Omega}
\end{figure}

\textit{Results.--}
All results obtained below are for $l=m=2$ $r$-modes, 
since the $l=m=2$ nodeless normal $r$-mode is believed to be the most unstable one \cite{ak01}.
Figure \ref{Fig:sigma1Omega} shows how $\sigma_1$ depends on 
$\nu$
for an NS with the mass $M=1.4M_\odot$ and
$T^\infty=10^7\,\rm K$, 
assuming BSk24 EOS, and SF model I. 
In this case the
$r$-mode spectrum consists of one normal nodeless $r$-mode $n_0$, 
one SF nodeless $r$-mode $s_0$, 
and an infinite set of SF modes with nodes 
(only the two first overtones, $s_1$ and $s_2$, 
are plotted in Fig.\ \ref{Fig:sigma1Omega}). 
Different modes are shown by different width lines.
Avoided crossings 
$n_0,s_1$ and $n_0,s_2$ are clearly visible where 
the modes
change their behavior from normal ($n_0$) to SF-like ($s_1$ or $s_2$) 
and vice versa. 
Dots 
show
the normal 
$r$-mode (at different temperatures different modes behave as the normal one). 

\begin{figure}
    \begin{center}
        \leavevmode
        \includegraphics[width=8.5cm]{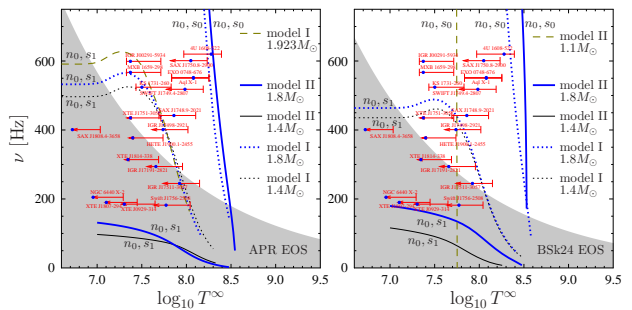}
    \end{center}
    \caption{The curves $\nu_{n_0,s_0}(T^\infty)$ (marked $n_0,s_0$) 
    	and $\nu_{n_0,s_1}(T^\infty)$ (marked $n_0,s_1$),
        showing $\nu$ and $T^\infty$ 
    	at which normal $r$-mode exhibits avoided crossings
    	with the $s_0$ and $s_1$ SF $r$-modes.
    	Left panel shows results for APR EOS, right panel -- for BSk24 EOS. 
Thick and thin curves 
are plotted for NSs with
$M=1.8M_\odot$ 
    	and 
    	$1.4M_\odot$, respectively. 
    	Dots correspond to model I, solid lines -- to model II. 
		Dashes in the left panel show $\nu_{n_0,s_1}(T^\infty)$
		for the maximum-mass configuration ($M=1.923M_\odot$) and model I.
		Dashes in the right panel show $\nu_{n_0,s_0}(T^\infty)$ 
		for $M=1.1M_\odot$ and model II. 
				}
    \label{Fig:inst}
\end{figure}

Since 
$Y_{ik}$ 
(and hence parameters $h$ and $h_1$, entering the oscillation equations) 
depends on $T^\infty$, 
the spin frequencies $\nu_{n_0,s_\alpha}$, 
at which avoided crossings $n_0,s_\alpha$ occur,
will also be temperature dependent 
($\alpha=0,1,2,$ and so on).
Fig.\ \ref{Fig:inst} 
shows the curves $\nu_{n_0,s_\alpha}(T^\infty)$ plotted for NSs 
with different masses,
assuming APR and BSk24 EOSs and models I and II of 
the neutron SF.
The gray area
is a classical stability region determined 
solely by the shear viscosity.
It is calculated for an NS with the mass $M=1.8M_\odot$.
The points correspond to available observational data on NSs in LMXBs \cite{gck14b,pw17}. 
The error bars describe uncertainties in the stellar envelope composition \cite{gck14b}. 
As one can see, many sources lie in the classical instability window (white region in Fig.\ \ref{Fig:inst}), 
which imposes a problem for a classical $r$-mode scenario. 
According to our proposal \cite{gck14a,gck14b},
the $r$-mode instability for these sources is suppressed because 
they are all located near the curves $\nu_{n_0,s_\alpha}(T^\infty)$, 
where $r$-mode dissipation is enhanced 
by
resonance interaction with one of the SF modes.

Generally (at not too high $T^\infty$; see below) 
avoided crossing $n_0,s_0$ 
takes place at an unrealistically high $\nu$. 
Only when 
$T^\infty$ 
approaches $T_{{{\rm c}n\,\rm max}}^\infty$
(the maximum of the redshifted $T_{{\rm c}n}$ 
in the region of the core, where neutron and proton SFs co-exist),
$\nu_{n_0,s_0}$
does start to decrease rapidly with increasing $T^\infty$ and vanishes at 
$T^\infty=T_{{{\rm c}n\,\rm max}}^\infty$. 
While $T_{{{\rm c}n\,\rm max}}=6\times 10^8\,\rm K$ 
for our SF models, 
$T_{{{\rm c}n\,\rm max}}^\infty$ depends on the NS mass 
and EOS through
the redshift parameter and varies in the range 
$T_{{{\rm c}n\,\rm max}}^\infty\sim (2.5-4)\times 10^8\,\rm K$. 
As a result, we have an almost vertical drop 
of $\nu_{n_0,s_0}$ 
at $T^\infty\sim (2-4)\times 10^8\,\rm{K}$ (see Fig.\ \ref{Fig:inst}).
(We do not plot $\nu_{n_0,s_0}$ for $M=1.4M_\odot$ NS 
because it is 
similar to the $M=1.8M_\odot$ case.)
The exception is low-mass 
configurations 
for the SF model II
(dashes on the right panel of Fig.\ \ref{Fig:inst}
corresponding to $M=1.1M_\odot$), 
for which $T_{{{\rm c}n\,\rm max}}^\infty$ is small 
(see Fig.\ \ref{Fig:Tc}). 

Avoided crossing 
$n_0,s_1$
lies at a lower $\nu$ than 
$n_0,s_0$.
For model I 
$\nu_{n_0,s_1}$
passes through the sources in the instability window and thus explains them.
At the same time, 
for model II, the corresponding 
$\nu_{n_0,s_1}$ is much lower.
This happens because in this model $T_{{\rm c}n}$ 
drops sharply in the outer core, 
shrinking the
SF region even at low $T^\infty$. 
As we checked for various SF models, such 
shrinking 
of the SF region 
in the outer core
leads to a dramatic decrease of 
$\nu_{n_0,s_1}$
at a given $T^\infty$. 
An analogous shrinking of the SF region due to a drop of $T_{{\rm c}n}$ at its higher-density slope 
(in the stellar center) 
also leads to a $\nu_{n_0,s_1}$ decrease;
this decrease, however, is not so pronounced.

\begin{figure}
    \begin{center}
        \leavevmode
        \includegraphics[width=6cm]{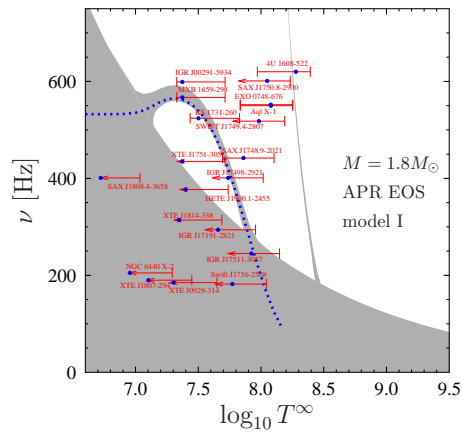}
    \end{center}
    \caption{Instability window for $l=m=2$ normal $r$-mode calculated for $M=1.8M_\odot$ NS with 
    	APR EOS and SF model I. In the filled region, the NS is stable. 
		Dotted line shows $\nu_{n_0,s_1}(T^\infty)$ 
		(the same line as in Fig.\ \ref{Fig:inst}). 
		We do not plot the curve $\nu_{n_0,s_0}(T^\infty)$ here,
		but it follows the corresponding stability peak.
   		}
    \label{Fig:peaks}
\end{figure}

Enhanced dissipation of the normal $r$-mode near avoided crossings stabilizes the mode,
and thus 
should affect 
the classical instability window.
This is illustrated in Fig.\ \ref{Fig:peaks}, 
which shows the instability window 
for an NS with $M=1.8M_\odot$, assuming 
APR EOS and SF model I. 
To calculate the window, 
we accounted for shear viscosity and mutual friction dissipation, 
as described in 
\cite{kg17}.
One notices the appearance of ``stability strips''
along the curves $\nu_{n_0,s_\alpha}(T^\infty)$
(they were termed stability peaks in the initial 
scenario of 
\cite{gck14a,gck14b}).
All the observed sources should be located 
in the 
stability region, which, however, varies with the stellar mass.

\textit{Discussion.--} We show that resonance stabilization of normal $r$-mode 
indeed occurs in the range of parameters relevant to NSs, 
falling 
within the classical instability window. 
In that sense, our calculations confirm the phenomenological scenario \cite{gck14a,gck14b} 
and put it on a solid ground.
Note that our results imply that stability peaks are not vertical (see Fig.\ \ref{Fig:peaks}), 
as the simple model \cite{gck14a,gck14b} 
predicted.
Nevertheless, an NS in the course of its evolution in LMXB will spend most 
of the time 
climbing up the left edge of the ``peak'', 
as  demonstrated in \cite{gck14a,gck14b}.

The presented calculations allow us to 
constrain models of a neutron SF. 
(Note that the proton SF only weakly affects the oscillation modes \cite{ga06,gkcg13,gkgc14}
and thus cannot be constrained by observations.)
Namely,
hottest rapidly rotating sources can be stabilized 
by resonance interaction of normal $r$-mode $n_0$ with the main harmonic $s_0$ of the SF $r$-mode.
Since the curve $\nu_{n_0,s_0}(T^\infty)$ 
for this resonance falls almost vertically 
to zero at $T^\infty=T_{{{\rm c}n\,\rm max}}^\infty$, 
we can constrain the value of $T_{{{\rm c}n\,\rm max}}$:
it should be $T_{{{\rm c}n\,\rm max}}\sim (3-6)\times 10^8\,\rm K$. 
The $T_{{\rm c}n}$ profiles adopted in this paper correspond 
to the upper limit of this range, 
and Fig.\ \ref{Fig:inst} implies that higher values of $T_{{{\rm c}n\,\rm max}}$ are not favorable. 
On the other hand, 
changing $T_{{{\rm c}n\,\rm max}}$ to the value $3\times 10^8\,\rm K$ 
shifts $\nu_{n_0,s_0}(T^\infty)$ to the left end of the error bar 
for
the hottest source 4U 1608-522, so that further decrease of $T_{{{\rm c}n\,\rm max}}$ 
complicates interpretation of this source.
We should stress that real maximum of $T_{{{\rm c}n}}$ can be larger than $T_{{{\rm c}n\,\rm max}}$ (the maximum of $T_{{\rm c}n}$ 
in the region of the core, where neutron and proton SFs co-exist). 
Thus, our estimate $(3-6)\times 10^8\,\rm K$ 
is the {\it lower limit} for the maximum value of $T_{{{\rm c}n}}$.
This estimate is in line with 
the constraints following from observations of cooling NSs 
\cite{gkyg04,plps04,gkyg05,syhhp11,page11,CasA13,CasA15,bhsp18} 
and it does
not contradict microscopic calculations \cite{ls01, yls99,gps14,dlz14,drddwcp16, sc19}. 
NSs in the instability window that are
not too hot 
may be stabilized by the resonance $n_0,s_1$
for some NS models, but only 
if $T_{{\rm c}n}$ profile is sufficiently wide, 
which ensures that the 
neutrons are superfluid in a
significant part of the NS core
at temperatures relevant to NSs in LMXBs.
An example of a wide profile is our model I.
Otherwise, if $T_{{\rm c}n}$ profile is narrow, 
like in our model II, 
explanation of 
moderately heated 
sources is 
problematic.
In such cases,
they may be stabilized only by the resonance with the main harmonic of the SF $r$-mode 
if we assume low masses for these sources (see the dashes in the right panel of Fig.\ \ref{Fig:inst}). 
But this 
alternative
is questionable
since NSs in LMXBs are believed to 
have high masses 
\cite{ozel12,antoniadis16}. 

An analysis of Fig.\ \ref{Fig:inst} shows that the interpretation of the existing sources
depends not only on the SF model but also on the EOS. 
While both EOSs employed in this Letter
are considered to be realistic, 
they lead to substantially different (by tens of percent) 
curves 
$\nu_{n_0,s_\alpha}(T^\infty)$
(compare the two panels in Fig.\ \ref{Fig:inst}). 
This occurrence opens up an 
intriguing
possibility 
of using observations of NSs in LMXBs not only for constraining 
SF models but also for constraining 
EOS of superdense matter, 
which 
is 
still poorly known at high densities \cite{ohkt17,bf18}.
To reach this goal, more accurate calculations of $r$-mode spectrum are required,
accounting for General Relativity effects, gravitational field perturbations,
and 
higher-order terms in the expansion (\ref{expan}).

In this Letter we 
focus
on
the resonance stabilization scenario,
which we consider as a minimal extension of the classical scenario, 
capable of explaining observations.
However,
other mechanisms of $r$-mode stabilization 
should 
also 
operate
in NSs,
first of all,
Ekman layer dissipation
\cite{bu00,lu01,glama06},
as well as
bulk viscosity in hyperon/quark matter
\cite{no06,alford12,oghf19},
enhanced mutual friction dissipation \cite{hap09} etc.
Real instability window may be a result of interplay of various stabilization mechanisms.

\begin{acknowledgments}
We thank A.I.~Chugunov for discussion.
MG acknowledges the financial support from the Foundation for the Advancement of Theoretical Physics 
and Mathematics BASIS (Grant No.\ 17-12-204-1) and from RFBR (Grant No.\ 18-32-20170). 
EK and VD thank Russian Science Foundation (Grant No.\ 19-12-00133) 
for the financial support of spectra calculations.
\end{acknowledgments}


\end{document}